\documentclass{fairmeta}

\usepackage{amsmath}
\usepackage{amsfonts}
\usepackage{amssymb}

\usepackage{enumitem}

\usepackage{graphicx}
\usepackage{url}

\graphicspath{{./}}

\let\origtextcolor\textcolor
\renewcommand{\textcolor}[2]{%
  \def\temp{#1}%
  \def\blue{blue}%
  \ifx\temp\blue
    #2%
  \else
    \origtextcolor{#1}{#2}%
  \fi
}

\setcitestyle{numbers,square,comma}

\title{OCC: Physical-Layer Assisted Congestion Control for Real-Time Communications}

\author[1]{Yufan Zhuang}
\author[1]{Zili Meng}
\author[1]{Zehong Lin}
\author[1]{Jun Zhang}
\affiliation[1]{Department of Electronic and Computer Engineering, The Hong Kong University of Science and Technology}

\metadata[Correspondence to]{Zehong Lin (\email{eezhlin@ust.hk}) and Jun Zhang (\email{eejzhang@ust.hk})}

\abstract{
Real-time communications (RTC) is a core technology for emerging applications in 6G, such as cloud gaming, teleoperation, and extended reality (XR), which require consistently low latency and high bitrates.
Existing RTC solutions fundamentally struggle to maintain low latency while supporting high bitrates due to their reliance on \textit{trial-and-error}-based mechanisms.
These mechanisms fail to probe the available bandwidth (ABW) promptly and accurately, leading to a trade-off between latency reliability and bandwidth utilization.
The tension becomes extremely more critical as the cellular bandwidth and application's demand fluctuate with a larger range in cellular networks nowadays.
To address this trade-off, we propose OCC, a novel approach that utilizes physical-layer information to \textit{explicitly obtain the ABW in real time}, enabling rapid adaptation to dynamic wireless network conditions.
However, the unique characteristics of RTC, including traffic bursts, application (APP) limits, and encoder lag, make the physical-layer informed control non-trivial.
OCC effectively addresses these issues through three innovative strategies: frame-aware bandwidth measurement, APP-limit-aware bandwidth estimation, and encoder-friendly rate control.
Extensive over-the-air experiments on an open-source cellular testbed demonstrate that OCC significantly enhances the performance of mobile RTC, reducing tail network latency by $13\%$ to $68\%$ and improving video frame bitrate by $1.2\times$ to $3.5\times$.

}

\begin{document}

\maketitle

\section{Introduction}
Real-time communications (RTC) serves as a fundamental technology for immersive applications in 6G networks, including cloud gaming, extended reality (XR), and teleoperation.
These applications demand ultra-responsive interactions and high-quality media delivery to provide a seamless and engaging user experience.
Meanwhile, there is a growing trend toward accessing RTC services over cellular networks, driven by increasingly affordable service plans and the expectation of ubiquitous and seamless connectivity \cite{mobile, ericssonreport}.
While 5G technology has achieved an average one-way network latency of $21.8$ ms, it still suffers from significant jitter \cite{xu2020understanding}.
To enable truly immersive RTC applications, it is essential to maintain consistently ultra-low latency
and remarkably high bitrates.
For instance, XR applications require latency under $60$ ms and bitrates exceeding $100$ Mbps \cite{fang2025immersive, de2024ten}, much higher than traditional video chat applications.

However, existing solutions struggle to support high bitrates while ensuring low latency.
Traditional rate control algorithms, such as congestion control (CC) \cite{carlucci2017congestion, arun2018copa, wang2024pudica} and forward error correction (FEC) \cite{li2023art, rudow2023tambur, antooth}, are typically \textit{conservative}.
For example, google congestion control (GCC) \cite{carlucci2017congestion} fundamentally struggles to support bitrates beyond $30$ Mbps (see Section \ref{sec2.2}) and underestimates the available bandwidth (ABW) in over $90\%$ of the time \cite{huang2025ace}, while FEC even consumes the bandwidth and limits effective throughput.
This conservative behavior stems from their reliance on a \textit{trial-and-error} mechanism for rate control, which suffers from a trade-off between latency and throughput.
In the face of network uncertainty, where the ABW can fluctuate by up to $50\times$ within $100$ ms, these solutions often sacrifice video bitrates to maintain latency reliability.
Consequently, this conservative approach results in suboptimal bandwidth utilization.
The challenge is exacerbated by the rapidly increasing bitrate requirements, from hundreds of Kbps for video calls to over $100$ Mbps for XR, while the growth in mobile access bandwidth from 4G to 5G remains modest \cite{li2025four}, approaching the limits of channel capacity.
This fundamental tension calls for the utmost utilization of available physical-layer bandwidth all the time.

The key to fully utilizing the capacity is to use the physical-layer information to open the box of uncertainty in the network.
While transport-layer solutions have to conservatively guess the capacity with trial-and-error mechanisms, physical-layer information can \textit{explicitly} reveal the ABW and detect bandwidth fluctuations from hundreds of Kbps to 100 Mbps in real time \cite{xie2020pbe, park2018exll, goyal2020abc}.
Such instantaneous feedback from the physical layer enables RTC senders to precisely align their sending rates with the ABW, thereby avoiding overshooting and underutilization of the network.
Moreover, the recent emergence of open radio access network (O-RAN) architectures \cite{polese2023understanding, azariah2024survey} offers unprecedented flexibility to modify physical-layer components, facilitating the deployment of physically informed solutions in commercial mobile systems.
Existing in-network feedback mechanisms also enable rapid feedback to the sender without requiring protocol modifications~\cite{flores2016enabling, goyal2020abc}, further easing integration.

\begin{figure}[t]
    \centering
    \includegraphics[width=0.5\linewidth]{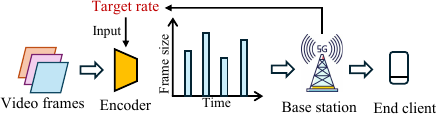}
    \caption{Overview of OCC. Physical-layer information helps the RTC sender know the ABW precisely and timely.}
    \label{overview}
\end{figure}

Motivated by this, we propose OCC, a novel physical-layer-informed control framework for RTC that achieves fine-grained rate control.
The design overview of OCC is shown in Fig. \ref{overview}.
Specifically, OCC monitors RAN metrics at the cellular base station (BS) and calculates the ABW for an RTC client.
Then, the result is fed back to the RTC sender as the target rate for the video encoder.
While the idea is intuitive, naively applying physical-layer solutions in real-world RTC systems leads to suboptimal performance due to the unique characteristics of RTC.
First, RTC traffic is typically bursty because of its frame-by-frame sending mechanism, which complicates ABW estimation and reduces accuracy to as low as $71\%$ (see Section \ref{traffic burst}).
Second, temporary traffic downturns caused by application (APP) limits lead to inadequate radio resource allocation, causing the physical layer to underestimate the ABW by up to $50\%$ (see Section \ref{app limit}).
Lastly, the encoding lag in video senders hinders the timeliness of physical-layer feedback, as video encoders may struggle to keep up with rapid changes in physical-layer rates, leading to overshooting and increasing tail latency to over $200$ ms (see Section \ref{encoder lag}).
To address these challenges, OCC introduces three innovative designs:
\textit{i) Frame-aware physical-layer measurement:} OCC employs a measurement approach that aligns bandwidth estimation with video frame intervals to mitigate the effects of traffic burstiness and enhance stability;
\textit{ii) APP-limit-aware bandwidth estimation:} OCC incorporates a margin to the estimated bandwidth and actively probes for potential traffic increases in APP-limited flows to ensure a full utilization of bandwidth;
\textit{iii) Encoder-friendly rate control:} OCC smoothly adjusts the sending rate to allow the encoder to keep pace with target rate, preventing overshooting and reducing tail latency.
Moreover, OCC dynamically detects bottleneck conditions and reverts to end-to-end solutions (e.g., GCC \cite{carlucci2017congestion}) when the bottleneck does not occur in the wireless link, thereby maintaining latency consistency.

We implement OCC in an over-the-air environment using an O-RAN testbed as a proof-of-concept.
Evaluation results demonstrate that OCC reduces network latency by 13\% to 68\% while simultaneously improving video bitrate by $1.2\times$ to $3.5\times$, significantly enhancing the user experience.
Furthermore, comprehensive experiments in dynamic network conditions, including mixed workloads, bottleneck switches, and user mobility, validate the robustness and practicality of OCC in complex real-world network environments.

Our main contributions are summarized as follows:
\begin{itemize}
    \item We motivate the need for utilizing physical-layer information to address the latency-bitrate trade-off in mobile RTC and identify three specific RTC characteristics that complicate the design process.
    \item We propose OCC, which incorporates three innovative orthogonal designs to overcome the identified RTC-specific challenges in the utilization of physical layer information, with the aim of achieving fine-grained physical-layer-informed control.
    \item We present a proof-of-concept implementation of OCC on WebRTC \cite{webrtc} and an O-RAN testbed and provide extensive evaluations based on over-the-air experiments.
\end{itemize}

\section{Background and Motivation}\label{sec2}
\subsection{Background}\label{sec2.1}
RTC requires consistently low latency and high bitrates for seamless interaction and an immersive experience.
However, current solutions struggle to achieve both at the same time.
On one hand, RTC applications demand ultra-low latency to ensure a seamless and immersive user experience.
For example, as shown in Table \ref{tab1}, videoconferencing typically requires a network delay of $< 130$ ms \cite{jones2021internet}, while cloud gaming demands a latency of $< 96$ ms \cite{kamarainen2017measurement}.
The latency requirements become more stringent for XR applications, which require latency consistently below $60$ ms \cite{fang2025immersive, de2024ten}.
Moreover, there is an ongoing trend to emphasize \textit{tail latency} (i.e., the higher percentiles of latency distributions), which often leads to stalls in the system.
With a tail latency of the 99.9-th percentile, the video streaming will stall once every 1000 frames, usually no more than $30$ s.
This illustrates why even ``rare" latency spikes have a greater impact than the average latency during long sessions.

On the other hand, a sufficiently high video bitrate is crucial for user engagement in RTC applications, as it allows for enhanced realistic details and smoother animations.
The required bitrates for various types of videos are shown in Table \ref{tab2}, from $3-4$ Mbps for high-definition (HD) videos, $15-25$ Mbps for 4K videos, $25-40$ Mbps for 360$^\circ$ videos, to $100-600$ Mbps for 6-DoF videos \cite{zoom, Broadband}.
Moreover, the bitrate requirements for RTC scale much faster than the wireless bandwidth and are about to overwhelm the channel capacity \cite{meng2022achieving, li2025four}.
Thus, fully utilizing the wireless bandwidth is necessary for the sustainable development of RTC.

\begin{table}[t]
    \centering
    \caption{RTC latency requirements.}
    \label{tab1}
    \setlength{\tabcolsep}{4pt}
    \begin{tabular}{cccc}
        \toprule
        Videoconferencing & Tele-robotics & Cloud gaming & AR/VR \\
        \midrule
        $< 130$ ms & $< 100$ ms & $< 96$ ms & $< 60$ ms \\
        \bottomrule
    \end{tabular}
\end{table}

\begin{table}[t]
    \centering
    \caption{RTC bitrate requirements.}
    \label{tab2}
    \setlength{\tabcolsep}{4pt}
    \begin{tabular}{cccc}
        \toprule
        Full HD videos & 4K videos & 360$^\circ$ videos & 6-DoF videos \\
        \midrule
        3$-$4 Mbps & 15$-$25 Mbps & 25$-$40 Mbps & 100$-$600 Mbps \\
        \bottomrule
    \end{tabular}
\end{table}

Although numerous efforts in RTC have been made to provide consistently low latency, they struggle to support high bitrates for emerging RTC scenarios like 6-DoF video streaming.
The fundamental reason is their reliance on the \textit{trial-and-error} principle for rate control.
To avoid network congestion and enhance bandwidth utilization, RTC senders should carefully determine the sending rate.
Transport-layer solutions, such as GCC \cite{carlucci2017congestion}, Copa \cite{arun2018copa}, ExStream \cite{park2024exstream}, and Pudica \cite{wang2024pudica}, typically begin video streaming with a low bitrate and incrementally increase it until network congestion is detected (e.g., a positive delay gradient in GCC).
Upon detecting congestion, the bitrate is sharply reduced to prevent further congestion and drain the packets queued in the network.
End-network cooperative solutions (e.g., XCP \cite{katabi2002congestion}, ABC \cite{goyal2020abc}, and Zhuge \cite{meng2022achieving}) leverage in-network devices to shorten the feedback loop, but the network-layer metrics they rely on, such as queue states, still cannot fundamentally reveal the ABW.
As a result, these solutions also fall into the trial-and-error paradigm when estimating the actual bandwidth.
Application-layer solutions, such as FEC \cite{li2023art, rudow2023tambur, antooth} and adaptive codecs \cite{fouladi2018salsify, ray2019vantage, zhou2019learning}, highly depend on the network state information provided by the CC mechanisms.
For example, the code rate of FEC is determined based on instantaneous network conditions to balance error tolerance and bandwidth overhead, and adaptive codecs also rely on CC mechanisms towards fine-grained coding control.
Thus, these higher-layer solutions inherit the limitations of the trial-and-error principle.

\subsection{Motivation}\label{sec2.2}

\begin{figure}[t]
    \centering
    \begin{minipage}[t]{0.25\linewidth}
        \centering
        \includegraphics[width=\linewidth]{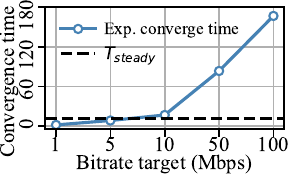}
        \caption{Expected bitrate converge time for GCC.}
        \label{fig29}
    \end{minipage}
    \hspace{1em}
    \begin{minipage}[t]{0.25\linewidth}
        \centering
        \includegraphics[width=\linewidth]{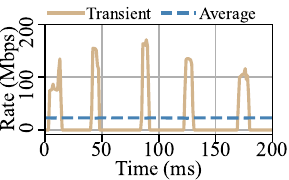}
        \caption{Illustration of RTC traffic bursts.}
        \label{fig19}
    \end{minipage}
    \hfill
\end{figure}

The trial-and-error mechanism fundamentally struggles to support high bitrates.
We illustrate this argument using GCC \cite{carlucci2017congestion} as an example.
GCC increases the sending rate $\bar{A}$ by the ratio between half of the packet size and the round-trip time (RTT) by default.
Even if there is no latency jitter at all, as shown in Fig. \ref{fig29}, it takes $83$ s to reach $50$ Mbps and $167$ s to reach $100$ Mbps (the blue curve).
This long duration far exceeds the maximum time $T_{steady}=11 \text{s}$ that the packet inter-arrival time can remain stable \cite{jansen2018performance} (the black dashed line).
Accelerating the rate increase, such as doubling $\bar{A}$, does not fundamentally solve this problem because it may increase the probability of overshooting, leading to more frequent bitrate reductions.
Thus, rate control solutions based on the conservative trial-and-error mechanism are limited by the trade-off between latency and throughput.
They struggle to effectively probe the ABW and fully utilize network resources, and consequently fall short of addressing the high bitrate requirements of emerging RTC applications.

The key to boosting the channel capacity for RTC applications is to utilize physical-layer information to reveal the \textit{explicit} ABW instead of relying on trial-and-error-based probing.
This enables RTC senders to align their sending rates with instantaneous ABW, avoiding network congestion and enhancing bandwidth utilization.
To be more specific:

\textit{i) The physical layer reveals the ABW explicitly and accurately.}
The physical layer of the cellular BS allocates physical resource blocks (PRB) and chooses the modulation and coding scheme (MCS) for cellular users, which fundamentally determines the ABW.
The PRB allocation and MCS adjustment occur at intervals of 1 ms \cite{xie2020pbe}, which is much shorter than the sliding window sizes used in status-quo end-to-end solutions \cite{carlucci2017congestion, wang2024pudica}.
While traditional methods may require more than one RTT for reliable ABW measurement, physical-layer-informed solutions can reduce this latency by two orders of magnitude, significantly shortening the response time to bandwidth changes \cite{xie2020pbe}.

\textit{ii) The physical layer senses the bandwidth changes in the most timely manner.}
As the lowest layer in computer networks, the physical layer is closest to the actual data transmission.
Fluctuations in the data rate at this layer directly lead to various packet-level incidents at the network layer, such as packet queuing, which subsequently result in transport- or application-layer events like packet delay or frame delay.
In contrast to methods that monitor queue states at the network layer \cite{flores2016enabling, katabi2002congestion} or rely on packet-delay-based measurements above the transport layer \cite{carlucci2017congestion, arun2018copa}, the physical layer can detect bandwidth changes directly and promptly, providing a more immediate and accurate reflection of network conditions.

\section{Observations and Design Challenges} \label{sec3}

While utilizing the physical-layer information for RTC rate control may seem straightforward, implementing it in real-world scenarios is non-trivial.
We find that the bursty nature of RTC traffic and APP limits hinder the physical-layer information from accurately indicating the ABW (see Section \ref{traffic burst} and \ref{app limit}).
In addition, even with precise ABW information, the video encoder may still overshoot the bandwidth and cause congestion (see Section \ref{encoder lag}).
The shortcomings of existing physical-layer-informed solutions motivate us to seek for new designs for RTC (see Section \ref{sec4}).

\subsection{Traffic Burst}\label{traffic burst}
The burstiness of RTC traffic is an outstanding feature that differentiates it from other traffic types.
\begin{itemize}[topsep=2pt, itemsep=0pt, left=0pt]
    \item \textit{Universality.}
    Traffic burstiness is common for RTC senders and arises from the frame-based sending pattern \cite{huang2025ace}.
    Specifically, RTC traffic consists of video frames that are generated in real time.
    Once these frames are encoded, all data packets are sent immediately.
    Although some senders utilize packet pacing to mitigate the adverse effects of sudden large data injections, they often set high pacing rates to minimize end-to-end latency (e.g., $2 \times$ in Copa \cite{arun2018copa} and $2.5 \times$ in GCC \cite{carlucci2017congestion}).
    \item \textit{Severity.}
    The burstiness of RTC traffic is positively correlated with the Internet bandwidth and video bitrates.
    For example, a single video frame may consist of 25 packets, and the available Internet bandwidth typically exceeds 500 Mbps \cite{xu2020understanding, li2025four}.
    Thus, injecting the 25 packets immediately into the network with this bandwidth may cause the transient throughput to reach hundreds of Mbps in the first few milliseconds, followed by idle periods with no data transmission until the next frame.
    With increasing Internet bandwidth and video bitrates, the burstiness of RTC traffic will become more severe.
\end{itemize}

To illustrate the bursty traffic pattern in RTC, we measure the transient sending rate (red curve) and the sending rate averaged over a window size of $1$ s (blue curve) for an RTC sender.
A 200 ms segment of this trace is shown in Fig. \ref{fig19}.
The frame rate of the video is 25 frames per second (fps), with each frame consisting of 25 packets, and the Internet bandwidth is 600 Mbps.
As illustrated in Fig. \ref{fig19}, the instantaneous sending rate spikes to $150$ Mbps within the first $2$ ms of a frame interval, then drops to zero for the remainder of the interval.

\begin{figure}[t]
    \centering
    \begin{subfigure}[b]{0.25\linewidth}
        \centering
        \includegraphics[width=\linewidth]{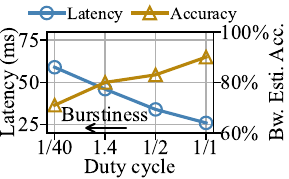}
        \caption{Effect of traffic bursts.}
        \label{fig1}
    \end{subfigure}
    \hspace{1em}
    \begin{subfigure}[b]{0.25\linewidth}
        \centering
        \includegraphics[width=\linewidth]{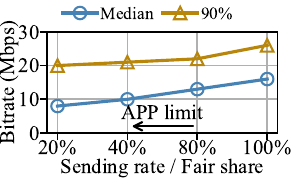}
        \caption{Effect of APP limits.}
        \label{fig2}
    \end{subfigure}
    \hfill
    \caption{(a) ABW estimation accuracy and related network latency under various levels of traffic bursts.
    We define the ``duty cycle'' as the fraction of a single frame interval during which data packets are transmitted.
    A smaller duty cycle indicates a more severe traffic burst.
    (b) The converged video bitrate when the sender is released from different levels of APP limits.
    We use the ratio of the actual sending rate to the fair-share bandwidth to describe the APP limit.
    A smaller ratio indicates a more significant APP limit.}
\end{figure}

\begin{figure}[t]
    \centering
    \includegraphics[width=0.45\linewidth]{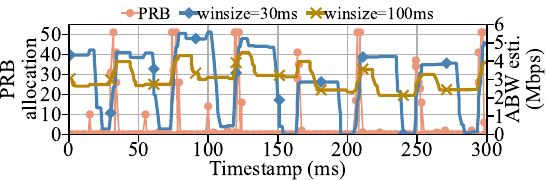}
    \caption{Illustration of unstable bandwidth measurements using moving averages with window sizes of 30 ms and 100 ms.
    Video frames are transmitted in bursts of 1 ms with a 40 ms interval between frames.}
    \label{fig7}
\end{figure}

\textbf{Finding 1: The bursty traffic pattern in RTC makes the physical-layer bandwidth measurement unstable.}
This is because the bursty traffic causes the physical-layer resource allocation to be bursty.
Therefore, existing physical-layer-based ABW estimation will have a fluctuating result depending on whether the burst is counted or not, as shown in the blue curve of the $30$ms-window measurement in Fig.~\ref{fig7}.
As a result, the estimated ABW deviates from the actual one, potentially leading to bandwidth overshooting and network congestion.

Moreover, this problem is exacerbated by the increase of traffic burstiness, since the more bursty the traffic, the greater the fluctuation in the physical-layer bandwidth estimation.
To demonstrate this, we measure the accuracy of the physical-layer bandwidth measurement and the network latency caused under different levels of traffic bursts (see Fig. \ref{fig1}).
We define ``duty cycle'' as the fraction of one frame interval during which data packets are transmitted, i.e., a duty cycle of $1/40$ means that the packets of one frame are bursted out in $1$ ms for a frame interval of $40$ ms.
Our measurement results show that, as the traffic burst becomes severe, the accuracy of the physical-layer bandwidth estimation (red curve) decreases from more than $90 \%$ to less than $40\%$, and the $99\%$-tail network latency (blue curve) increases from less than $30$ ms to around $60$ ms.

\subsection{APP Limit}\label{app limit}
The traffic of RTC may be limited by the application instead of the network, a phenomenon known as the APP limit.
This is because the video data for RTC are generated in real time with various bitrates due to changes in video content.
For example, in a lightweight first-person shooter game, the rapid movement of the player can cause the video bitrate to reach $\sim 2$ Mbps, while a static scene may only require $\sim 600$ Kbps. In this case, the video bitrate is limited by the application itself instead of the network bottleneck.
Moreover, the variance in video bitrates will continuously increase as streaming applications increasingly pursue higher fidelity, making the APP limit a common issue.

\textbf{Finding 2: The physical layer may underestimate the ABW for an APP-limited flow, leading to possible bandwidth underutilization and low bitrates.}
This is because the physical-layer ABW estimation is based on actual traffic.
When the APP-limited traffic is lower than the fair share, the BS can never know where the fair share is.

We vary the level of the APP limit and show that a more significant APP limit leads to lower bitrates for both the median (blue curve) and the 90-th percentile (red curve), as illustrated in Fig. \ref{fig2}.
This is because the BS allocates just enough radio resources to each flow, unless the total traffic demand exceeds the capacity \cite{xu2022tutti}.
Consequently, the estimated ABW at the physical layer may be lower than the actual ABW for an APP-limited flow, which causes insufficient bandwidth utilization and leads to low bitrates.

\subsection{Encoder Lag}\label{encoder lag}
The video encoder may still overshoot the bandwidth even with precise ABW information \cite{fouladi2018salsify}.
Similarly to other RTC rate control schemes, the estimated ABW in OCC is set as the target encoding bitrate for the video encoder.
However, the encoding bitrate cannot exactly follow the input target bitrate, due to the following two reasons:
\begin{itemize}[topsep=2pt, itemsep=0pt, left=0pt]
\item \textit{Coding efficiency}.
The majority of video codecs, such as H.264 or VP8, use predictive coding, where the encoder predicts the content of a frame based on previous frames.
Significant variability in target frame sizes can lead to inefficient compression.
\item \textit{Video quality}.
For RTC applications, maintaining a stable bitrate is crucial to ensure a smooth viewing experience.
Significant fluctuations in bitrate can lead to inconsistent quality across frames, degrading the user experience.
\end{itemize}

\begin{figure}[ht]
\centering
\begin{minipage}[t]{0.75\linewidth}
    \centering
    \includegraphics[width=\linewidth]{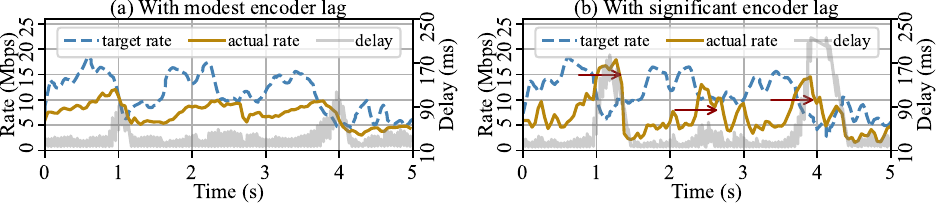}
    \caption{Illustration of the encoder lag, where the physical-layer rate is directly used as the target rate. We configure different VBV buffer sizes in the x264 encoder to simulate different degrees of encoder lag.}
    \label{fig3}
\end{minipage}
\hfill
\begin{minipage}[t]{0.22\linewidth}
    \centering
    \includegraphics[width=\linewidth]{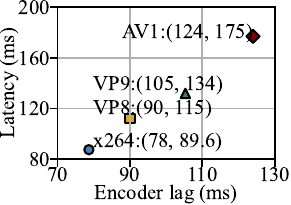}
    \caption{The encoder lag and related tail latency for the mainstream codecs.}
    \label{fig22}
\end{minipage}
\end{figure}

\textbf{Finding 3: The video encoder cannot catch up with the unsmoothed physical-layer rate, causing bandwidth overshooting even with a precise ABW estimation.}
The physical-layer measurement ABW is obtained directly from the real data channel without smoothing from inter-layer buffers, which makes it unsuitable for direct use as the target rate for the video encoder.
Consequently, the encoder may not strictly adhere to the target rate, and instead has some lag.

To illustrate the encoder lag, we record the behavior of the x264 video encoder when fed with raw physical-layer rates as target rates and monitor the corresponding network latency.
We configure different VBV buffer sizes in the x264 encoder to simulate different degrees of encoder lag.
As shown in Fig. \ref{fig3}, when the physical-layer rate (blue curve) increases from 10 Mbps to 20 Mbps and then sharply decreases to 5 Mbps, the encoder lags behind the bandwidth drop for approximately 350 ms while still increasing the sending rate (red curve).
This results in bandwidth overshooting and increased latency over 160 ms (gray curve).

More importantly, the congestion caused by the encoder is a common issue for most of the video codecs used today \cite{fouladi2018salsify, zhou2019learning, ray2022sqp}.
We measure the encoder lag (x-axis) and related tail network congestion (y-axis) for different codecs including x264, VP8, VP9, and AV1, as shown in Fig. \ref{fig22}.
The encoder lag is defined as the response time of the video encoder to a bandwidth drop, specifically the event when the ABW drops by more than 50\% within 100 ms.
The results show that the average lag for the four encoders to the bandwidth drop events is 103 ms, which is sufficient to congest the network.

\section{OCC Design} \label{sec4}
To address the design challenges identified in Section \ref{sec3}, OCC incorporates three orthogonal designs, with the aim of achieving fine-grained physical-layer-informed congestion control for RTC.
Specifically:
\begin{itemize}
    \item In order to eliminate the measurement noise from traffic bursts, OCC employs a frame-aware physical-layer measurement to achieve accurate bandwidth estimation (see Section \ref{sec4.1}).
    \item To accurately estimate resources for APP-limited flows, OCC implements an APP-limit-aware bandwidth estimation strategy and probes potential traffic increases (see Section \ref{sec4.2}).
    \item To reduce encoder overshoot, OCC smooths the physical-layer bandwidth measurement by taking the minimum over a specific past period (see Section \ref{sec4.3}).
\end{itemize}
Finally, we discuss how OCC switches between bottleneck states to ensure latency consistency and how to carry the ABW information back to the sender (see Section \ref{sec4.4}).

\textbf{\textit{Physical-layer bandwidth estimation.}}
The first step of OCC is to estimate the ABW from the physical layer.
OCC calculates the ABW according to the radio resource allocation and the MCS rate, as in \cite{xie2015pistream, xie2020pbe}.
Specifically, the physical-layer capacity $C_p$ for a cellular user is expressed as
\begin{equation} \label{Cp}
    C_p = (P_{allocated} + P_{idle} / N_{user}) \cdot R_{mcs},
\end{equation}
where $P_{alloacated}$ and $P_{idle}$ denote the number of allocated and idle PRBs, respectively, $N_{user}$ denotes the number of users in the cell, and $R_{mcs}$ denotes the MCS rate.
The capacity $C_p$ can then be converted to the transport-layer goodput $C_t$, which is regarded as the ABW estimation and subsequently fed back to the sender.
\footnote{For simplicity, we do not distinguish between $C_p$ and $C_t$ in the rest of this paper, instead referring to both as the physical-layer capacity $C_p$. A detailed explanation of the translation between $C_p$ and $C_t$ can be found in \cite{xie2020pbe}.}

\subsection{Frame-aware Measurement} \label{sec4.1}
The key to effective rate control for RTC is obtaining an accurate ABW estimation.
However, the bursty nature of RTC traffic leads to the fluctuation in the physical-layer capacity estimation (see Section \ref{traffic burst}).

\textbf{Insight 1: The denoising method for bandwidth measurement should match the temporal RTC traffic pattern.}
Applying a moving average over one RTT as in \cite{xie2020pbe} is suboptimal since the temporal characteristics of RTC flows are not related to RTT, as discussed in Section \ref{traffic burst}.
Moreover, averaging over a large window size eliminates the noise but inevitably leads to insensitivity to bandwidth drops.
Thus, to obtain an accurate and stable ABW estimation, we need to \textit{align with the temporal traffic patterns of RTC flows} in the bandwidth measurement from the physical layer, which has been proven to be effective in higher-layer solutions \cite{zhang2024habitus, park2024exstream}.

\begin{figure}[t]
    \centering
    \includegraphics[width=0.35\linewidth]{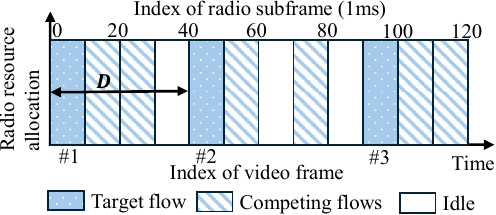}
    \caption{Illustration of frame-aware measurement. The figure shows the radio resource allocation for multiple flows. The target flow is an RTC flow with three video frames arriving at $t=0$ ms, $t=40$ ms, and $t=90$ ms, respectively. Each frame takes $10$ ms (i.e., 10 radio subframes) to be drained from the BS. During the remaining time, the resources are allocated to serve competing flows or left idle.}
    \label{frame_bw_esti}
\end{figure}

\textbf{Design 1: OCC utilizes a frame-aware measurement to estimate the ABW for an RTC flow.}
According to the frame structure in cellular networks, the transmission of one radio subframe occupies 1 ms and the radio resource allocation is adjusted at this level of granularity \cite{xie2020pbe}.
If the interval between two video frames is $D$ ms, there are $D$ radio subframes in between.
OCC estimates the ABW, denoted by $B$, during the frame interval as follows:
\begin{equation}\label{B1}
    B = \frac{1}{D} \times \sum_{i=1}^{D} C_{p,i},
\end{equation}
where $C_{p,i}$ denotes the physical-layer capacity estimated using (\ref{Cp}) in the $i$-th radio subframe.
By utilizing a video-frame-aware measurement, OCC achieves more fine-grained bandwidth estimation for RTC flows, distinguishing it from existing works (see Fig. \ref{frame_bw_esti}).

\textbf{\textit{Frame interval identification.}}
In RTC systems, the frame interval $D$ can vary among different RTC flows, and even within one RTC flow, $D$ is not fixed during the streaming.
It can change based on video content or user preferences.
In addition, the frame rate configuration is an application-layer parameter that is invisible in the L1/L2 BS.
Therefore, negotiating this information between the video sender and the BS requires additional protocols and signaling overhead.
To address this issue, a data-transparent solution is necessary for identifying the frame interval $D$ in the BS.

In response, OCC leverages the bursty traffic pattern of RTC to identify each video frame.
In the cellular BS, there are separate downlink (DL) buffers for each user, ensuring that traffic for different users remains isolated.
Since the packets of a single video frame are sent in bursts and the wireless hop is the bottleneck in most cases \cite{xie2017accelerating}, the packets arrive at the DL buffer in the BS as a packet train.
There is minimal data enqueueing before the next frame arrives, expect some negligible control traffic.
By identifying the start and end of these packet bursts in the DL buffer, OCC can dynamically recognize the duration $D$ of each video frame, thus enabling frame-aware ABW measurement.

\textbf{\textit{Fast feedback.}}
Although the frame-aware measurement enhances the stability and accuracy of bandwidth estimation for RTC flows, it may delay the ABW feedback, which is critical in dynamic wireless channels.
The frame interval of RTC is typically tens of milliseconds (e.g., $40$ ms for a $25$ fps video), so as the sample period of the bandwidth measurement.
However, the channel coherence time is shorter than this sample period.
This discrepancy means that sharp channel degradations (deep fading) can occur during the sample interval, but the sender may not be aware of these changes until the measurement for an entire video frame is completed.
This will result in a delayed response from the sender, leading to bandwidth overshooting and increased network latency.

To provide timely ABW information to the sender within a millisecond granularity, OCC measures the ABW based on video frames while considering real-time wireless channel conditions.
The event of channel degradation can be sensitively detected at the BS, enabling it to select a lower order of MCS for the next radio subframe.
Thus, OCC feeds back the ABW calculated by the MCS rate of the latest radio subframe and the PRB utilization measured from the last video frame.
Consequently, the ABW measurement in (\ref{B1}) is revised as:
\begin{equation}
    B = \frac{1}{D} \cdot \sum_{i=1}^{D} \left(P_{allocated,i}+\frac{P_{idle,i}}{N_{user}} \right) \cdot R_{mcs, now},
\end{equation}
where $R_{mcs, now}$ denotes the MCS rate of the latest radio subframe (the most recent 1 ms), $P_{allocated,i}$ and $P_{idle,i}$ denote the number of allocated and idle PRBs in the $i$-th radio subframe, respectively, and $D$ is the interval for the last video frame.
Then, the ABW $B$ is calculated every millisecond.

\textbf{\textit{Fitting in paced traffic.}}
In industrial practice, RTC systems typically employ sender-side pacing to smooth out bursts.
However, pacing rates are often set well above the average encoding bitrate (e.g., $2.5\times$ in GCC \cite{carlucci2017congestion}) to minimize per-frame queuing delay, so the traffic still exhibits a burst-then-idle pattern with a moderate duty cycle.
When pacing fully eliminates observable idle gaps between frames, OCC falls back to legacy physical-layer ABW measurement.
Specifically, we set a threshold $D_{th}$ for frame interval identification.
If no idle period exceeding $D_{th}$ is detected in the downlink buffer, OCC measures the ABW over a fixed window of length $D_{th}$, which is functionally equivalent to the approach in PBE-CC \cite{xie2020pbe}.
This ensures that OCC remains robust regardless of the sender's pacing configuration.

\subsection{APP-limit-aware Bandwidth Estimation} \label{sec4.2}
Since the physical-layer bandwidth estimation is based on resource allocation, temporary traffic downturns due to APP limits result in inadequate radio resource allocation, adversely affecting the physical-layer measurement (see Section \ref{app limit}).

\textbf{Insight 2: Potential traffic increases from APP-limited flows should be considered when calculating the ABW.}
The physical-layer-informed ABW estimation is based on radio resource allocation, but the allocated resources do not always represent the maximum ABW, especially when the target flow is in the APP limit.
It is possible that additional radio resources can be safely utilized, allowing for an increase in the sending rate even if the total radio resource utilization appears full.

\textbf{Design 2: OCC allocates a margin for ABW estimation on flows where PRB utilization is below the fair share.}

\textbf{\textit{The fair-share capacity.}}
The ABW is at least the ``fair-share'' capacity $C_f$, defined as the converged resource allocation for the target user under the active scheduling algorithm.
Since OCC operates at the BS, it has direct access to the scheduler type and per-user channel state information, enabling accurate computation of $C_f$ for each user.
For example, under round-robin scheduling, $C_f = P_{total}/N_{user}\cdot R_{mcs}$.
Under proportional fair scheduling, $C_f$ reflects the user's expected allocation based on its channel quality relative to other users in the cell.
This scheduler-aware computation ensures that $C_f$ does not overestimate the achievable capacity for cell-edge users.
Here, $P_{total}$ denotes the total number of PRBs available at the BS.
This is because the fair-share PRB utilization $(P_{total}/N_{user})$ can always be satisfied by the MAC scheduler at the cellular BS.
Thus, to facilitate an APP-limited flow in reclaiming its fair-share bandwidth upon exiting the APP limit, it is reasonable for CC to set $\max\{C_p, C_f\}$ as the target rate instead of directly using $C_p$ in (\ref{Cp}).

\textbf{\textit{Increasing the rate gently.}}
However, using $\max\{C_p, C_f\}$ as the target rate may cause the sending rate to abruptly increase from a relatively low APP-limited value to $C_f$, leaving no response time to other flows.
Therefore, a more gentle method is required to guide the sender out of the APP limit.

To realize this, OCC introduces a margin associated with a coefficient $\beta$, where $0<\beta < 1$ and is set to 0.1 in our implementation.
If the estimated physical-layer capacity $C_p$ is less than the fair-share $C_f$, the ABW fed back to the sender is adjusted by adding the margin, which is equal to a portion of the difference between $C_f$ and $C_p$, specifically an increment of $\beta \cdot(C_f-C_p)$.
The capacity estimation in (\ref{Cp}) is thus revised as follows to indicate the actual physical-layer capacity $C_p^\prime$:
\begin{equation}
    C_p^\prime = C_p + \beta \cdot \max\{C_f - C_p, 0\}.
    \label{eq:app_limit_margin}
\end{equation}

Importantly, Eq.~(\ref{eq:app_limit_margin}) does not directly set the target rate to $C_f$; when a flow is persistently APP-limited, it is fed back with $C_p$ plus a small margin (controlled by $\beta$), which avoids intentional resource reservation and waste.
The margin is only used to accelerate recovery when the APP limit disappears.

\textbf{\textit{Adaptation to various schedulers.}}
OCC is generalizable to different cellular scheduling algorithms, as the fair-share capacity $C_f$ is computed at the BS with full knowledge of the active scheduler and per-user channel conditions.
Moreover, the incremental probing via the coefficient $\beta$ (Eq.~(\ref{eq:app_limit_margin})) ensures that even under estimation inaccuracies, the sender approaches $C_f$ gradually, preventing abrupt rate increases that could cause RLC/MAC-layer congestion.
If the probed rate exceeds the actual schedulable capacity, the physical-layer measurement $C_p$ in the subsequent frame interval will reflect the deficit, naturally correcting the target rate downward.

\subsection{Encoder-friendly Rate Smoothing} \label{sec4.3}
After obtaining a reliable ABW measurement, the final step of OCC is to set a target rate for the video encoder.
However, simply reflecting the instantaneous physical-layer rate can cause overshoot due to encoder lag, as discussed in Section \ref{encoder lag}.

\textbf{Insight 3: The standard deviation of the target rate needs to be bounded.}
The design goal of OCC is to ensure that the bitrate does not fluctuate by more than a certain percentage over a period, specifically, to control the bitrate standard deviation to below 2 Mbps.
This stability is crucial because excessive fluctuations hinder the encoder's ability to adapt efficiently, leading to potential performance degradation.

\textbf{Design 3: OCC takes the minimum of ABW estimations over a past period as the target rate for the video encoder.}
The challenge lies in the need to provide timely ABW feedback with millisecond granularity while avoiding the risk of encoder overshoot from an unsmoothed rate.
To address this dilemma and achieve an upper-bounded bitrate standard deviation, we have the following two arguments:

\textit{i) Applying a moving average is suboptimal.}
While applying a moving average for smoothing can meet the requirement of bitrate standard deviation, it may make the system insensitive to sharp bandwidth drops.
Consequently, significant decreases in bandwidth are averaged out, leading to potential overshoot.

\textit{ii) Ephemeral ABW increases should be bypassed}.
Ephemeral increases in the ABW can be utilized for temporary increases in frame bitrate, but these fluctuations should not necessarily be reflected in the target rate of the video encoder.
As discussed in Section \ref{encoder lag}, the video encoder struggles to support transient rate increases.
Moreover, a short-term rate increase is unlikely to enhance visual perception and may instead degrade quality consistency by introducing noticeable fluctuations.
Therefore, these ephemeral ABW increases should be bypassed.

Based on these arguments, OCC sets the target rate $R$ as the minimum observed physical-layer capacity $C_p$ over a sliding window of length $L$, where the window advances every 1 ms. The minimum operator ensures that any capacity drop is captured within a single subframe, while the window length determines how long this conservative estimate is retained as a safeguard.
Let $C_{p,l}$ denote the instantaneous bandwidth measurement in the $l$-th radio subframe. The target rate $R$ is calculated as follows:
\begin{equation}
    R = \min_{l\in [l-L,l]} C_{p,l},
\end{equation}
where $L$ is an adjustable parameter and is set to 500 in our implementation.
This strategy of taking the minimum over a sliding window enhances smoothness while maintaining sensitivity to changes in bandwidth.
Specifically, when a deep fade occurs, the minimum capacity within the window is immediately updated, allowing for timely detection of the bandwidth drop.
In contrast, any ephemeral ABW increase is filtered out by the window, which prevents the encoder from struggling with a fluctuating target rate.

When a deep fade occurs, the target rate may be fixed at a small value (e.g., 1/10 of the original rate) for the duration of the window length $L$. This is a side effect of the window-based design.
However, this limitation does not contradict our main objective of utilizing physical-layer information for rapid adaptation. End-to-end solutions typically take more than $100$ seconds to increase the sending rate by ten times, which is significantly longer than the window length $L$.

\subsection{Discussions} \label{sec4.4}

\begin{figure}[t]
    \centering
    \includegraphics[width=0.4\linewidth]{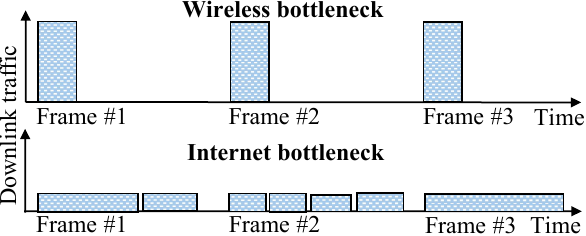}
    \caption{Traffic patterns observed at the BS under different bottleneck states.
    When the bottleneck is in the wireless link, the packets arrive at the BS in bursts, otherwise the packets arrive in a more continuous manner.}
    \label{in_out_time}
\end{figure}

\textbf{\textit{Bottleneck state switching.}}
In most cases, the network bottleneck lies in the wireless link \cite{meng2022achieving}.
However, when the bottleneck shifts to the Internet, OCC reverts to using the end-to-end solution GCC to maintain latency consistency.
Specifically, OCC identifies the bottleneck state at the BS by monitoring traffic patterns.
As noted in Section \ref{traffic burst}, RTC sends packets in bursts.
This bursty pattern will persist at the BS if the bottleneck is not the Internet, and vice versa (see Fig. \ref{in_out_time}).
Consequently, the presence or absence of burstiness at the BS serves as an indicator of the bottleneck states.

It is important to note that, in most common scenarios, there is a significant difference between wireless and wireline bottlenecks, especially in the presence of channel variations and cross-traffic disturbances \cite{xie2017accelerating}.
Therefore, even a coarse check is sufficient.
Moreover, since GCC makes decisions based on historical states, we continue to update the current network states to GCC while using OCC, ensuring that GCC can take over at any time.

\textbf{\textit{Feedback mechanism.}}
In terms of deployment, OCC requires coordination between the video sender and the BS, which follows the mechanism of previous end-network solutions \cite{katabi2002congestion, tai2008making, flores2016enabling, goyal2020abc, xie2020pbe, meng2022achieving}.
Specifically, the sender marks packets via header fields to enable BS identification of RTC flows.
Moreover, the BS utilizes the ECN bits in ACK packets \cite{goyal2020abc}, the payload of RTCP feedback messages \cite{meng2022achieving}, or a dedicated connection\cite{xie2020pbe}, to indicate the target bitrate to the sender.
These approaches are compatible with the existing Internet protocol.
In addition, the deployment of OCC is facilitated by the advent of O-RAN architectures \cite{polese2023understanding, azariah2024survey} that enable flexible software updates to the cellular firmware.

\textbf{\textit{Applicability to UL scenarios.}}
Although we mainly use the DL scenario as an illustration, OCC is directly applicable to uplink (UL) cases.
Specifically, the UL ABW can be computed from the same physical-layer information (PRB allocation and MCS), as the BS performs UL scheduling analogously to DL scheduling.
The UE can obtain its UL scheduling grant by decoding the physical downlink control channel (PDCCH) \cite{xie2020pbe}, enabling local ABW computation without explicit BS feedback.
Moreover, the three RTC-tailored designs (Designs 1-3) address properties of the RTC application layer, i.e., frame-level traffic bursts, APP-limited behavior, and encoder lag, that are independent of the link direction and thus apply to UL senders.

\textbf{\textit{Concurrent flows.}}
OCC's frame-aware measurement assumes that the target RTC flow dominates the UE's downlink traffic, which holds in most practical scenarios.
When lightweight background flows coexist (such as text messaging, push notifications, and heartbeat signals), their negligible volume does not distort the frame-level burst pattern (e.g., a few Kbps vs. tens of Mbps for the RTC stream).
The operating system on the end device typically deprioritizes bandwidth-intensive but latency-insensitive flows, such as file downloads, when there is an active RTC session.
Moreover, scenarios involving multiple concurrent RTC applications from the same UE are relatively rare in practice, because a user typically engages in one video call or one cloud gaming session at a time.
Thus, the assumption of a dominant RTC flow is reasonable for the target use cases of OCC.

\textbf{\textit{Handover.}}
In real cellular deployments, a UE may hand over between BSs.
OCC is inherently compatible with handover because the physical-layer capacity is locally available at the serving BS and does not depend on historical state from the previous BS.
During the brief handover transition, the BS detects the event via the standard radio resource control (RRC) handover procedure and signals the sender to fall back to GCC, to ensure continued functionality of RTC session without physical-layer feedback.
After the UE is attached to the new serving BS, the ABW estimation can be resumed immediately if the new BS supports OCC.

\begin{figure}[t]
    \centering
    \includegraphics[width=0.6\linewidth]{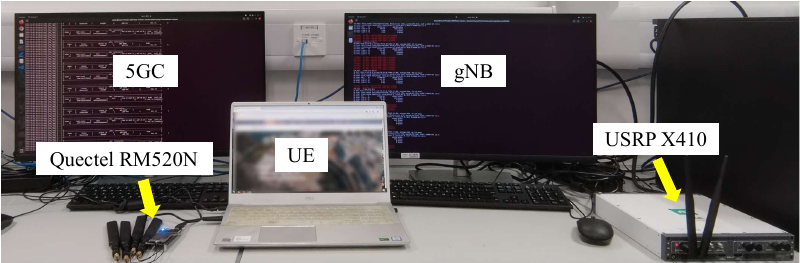}
    \caption{Physical layout of the O-RAN testbed.}
    \label{physical layout}
\end{figure}

\textbf{\textit{Parameter settings.}}
OCC involves three parameters: the frame interval detection threshold $D_{th}$ in Design~1, the probing coefficient $\beta$ in Design~2, and the sliding-window length $L$ in Design~3.
Specifically, $D_{th}$ is set based on the typical frame interval of the video encoder,
$\beta$ is assigned a small value (0.1) to enable gentle probing toward the fair-share capacity without overshooting, and
$L$ is set to 500~ms to filter out ephemeral bandwidth spikes that the encoder cannot effectively exploit.
All parameters are fixed across all experiments and do not require per-scenario tuning.

\section{Implementation} \label{sec5}

\begin{table}[t]
    \centering
    \caption{O-RAN testbed setup.}
    \label{tab3}
    \setlength{\tabcolsep}{6pt}
    \begin{tabular}{ll}
        \toprule
        \textbf{Radio Unit} & USRP X410 \cite{usrpx410}; 20 MHz at 3.3 GHz \\
        \textbf{User Equipment(s)} & Quectel RM520N-GL 5G \cite{quectelrm520n} \\
        \midrule
        \textbf{RAN Server} & Intel(R) Core (TM) i9-10980 \\
        \textbf{Core Network Server} & Intel(R) Core (TM) i9-10980 \\
        \midrule
        \textbf{5G RAN stack} & srsRAN \cite{srsRAN2022} \\
        \textbf{5G core} & open5GS \cite{open5gs2022} \\
        \textbf{OS kernel} & Linux kernel v5.4.0-26-generic \\
        \bottomrule
    \end{tabular}
\end{table}

We implement OCC on WebRTC \cite{webrtc}, the most widely-used RTC platform, and deploy it on an O-RAN testbed.
The server streams a timestamped video to the client via RTP/RTCP.
We utilize libx264 as the video codec.

\textbf{\textit{OCC workflow.}}
In terms of the control path, OCC monitors the RAN metrics at the physical layer of the O-RAN BS to estimate the ABW.
This ABW is then fed back as the target rate to the video encoder in the RTC sender.
For the data path, the RTC sender encodes video frames according to the target rate provided by OCC and streams these frames to the RTC receiver.
The RTC traffic travels through the Internet to the O-RAN cellular network and ultimately reaches the end device via the wireless link.

\textbf{\textit{O-RAN testbed.}}
We implement an O-RAN testbed to evaluate OCC in real cellular systems, leveraging the flexibility and openness of O-RAN architectures \cite{polese2023understanding, azariah2024survey}.
Programming a cellular BS typically requires customization of its firmware, but the source code for the current cellular firmware is proprietary to equipment manufacturers and thus inaccessible.
To address this challenge, we develop an open-source prototyping platform that utilizes O-RAN principles to replicate the full functionality of a cellular network.
This platform supports monitoring and analysis of RAN metrics without requiring any firmware customization.

The setup is summarized in Table \ref{tab3}.
The testbed emulates a 5G NR stand-alone (SA) system, consisting of a core network (5GC), a 5G BS (gNB), and several user equipments (UEs).
A physical layout is illustrated in Fig. \ref{physical layout}.
The 5GC and gNB are implemented using the open-source software srsRAN \cite{srsRAN2022} and open5GS \cite{open5gs2022}, respectively.
The software defined radio (SDR) USRP X410 \cite{usrpx410} serves as the radio front end for the gNB.
The UE is implemented using a Quectel RM520N-GL wireless modem module \cite{quectelrm520n} and a programmable SIM card, connected to a laptop as the host machine.
The cell is configured with a 20 MHz bandwidth with 51 PRBs, achieving a DL throughput of $\sim 50$ Mbps and an UL throughput of $\sim 15$ Mbps.

Importantly, despite involving lower network layers (e.g., the physical layer), our implementation relies only on open-source software and general-purpose hardware, making our work reproducible for the research community.

\section{Evaluation}\label{sec6}

\begin{figure}[t]
    \centering
    \includegraphics[width=0.4\linewidth]{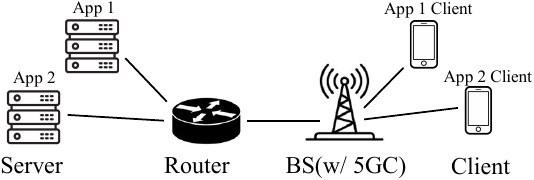}
    \caption{Experiment setup.
    Traffic flows are sourced from application services (servers) transmitting data to receivers (clients) through Ethernet links with a maximum capacity of $600$ Mbps and a cellular wireless link with a maximum capacity of $50$ Mbps.}
    \label{topology}
\end{figure}

We first introduce the experimental setup in Section \ref{sec6.1} and then evaluate the performance of OCC to answer the following questions:
\begin{itemize}
    \item \textit{Can OCC reduce latency without sacrificing or even improving bitrates in real systems?}
    Our prototype deployment of OCC shows that OCC reduces the network latency by 13\% to 68\%, while improving the video frame bitrate by $1.2\times$ to $3.5\times$. (see Section \ref{sec6.2})
    \item \textit{Can OCC effectively address the challenges posed by the three characteristics of RTC?}
    We craft scenarios with varying levels of traffic bursts, APP limits, and encoder lag.
    We observe performance improvements with OCC in most scenarios. (see Section \ref{sec6.3})
    \item \textit{How does OCC perform under dynamic conditions, including mixed workloads, bottleneck switching, and user mobility?}
    We demonstrate the decisions of OCC and relevant network metrics, showing that OCC is capable of adapting to fast-changing environments. (see Section \ref{sec6.4})
    \item \textit{What about the convergence speed, CPU overhead, and internal/external fairness of OCC?}
    We find that OCC converges fast to the target rate, ensures fairness with other flows, and incurs acceptable overhead. (see Section \ref{sec6.5})
\end{itemize}

\subsection{Experimental Setup} \label{sec6.1}
We follow the S-A-B-C setup, where S is a video server, A is a router, B is a 5G core and BS, and C is a video client, as shown in Fig. \ref{topology}.
In the following, we present the datasets, baselines, metrics, and physical environments used in the experiments.

\textbf{\textit{Datasets.}}
We use videos from the standardized UGC dataset \cite{wang2019youtube} for evaluation.
To demonstrate that OCC is scalable across various video content types, the test videos include 2K HDR videos with an average bitrate of 13.9 Mbps and 1080p lecture videos with an average bitrate of 7.6 Mbps, all at a frame rate of 25 fps.
The videos are transformed into raw videos and encoded in real-time during the streaming.

\textbf{\textit{Baselines.}}
We compare OCC with the following baselines:
\begin{itemize}
    \item \textit{GCC} \cite{carlucci2017congestion}, which is the state-of-the-art end-to-end congestion control algorithm for RTC.
    GCC achieves low latency in many experiments \cite{meng2022achieving, wang2024pudica} and is widely deployed in real streaming services like WebRTC \cite{webrtc}.
    \item \textit{PBE-CC} \cite{xie2020pbe}, which is the state-of-the-art physical-layer-informed congestion control solution.
    It integrates physical-layer measurements with the transport layer to enhance congestion control algorithms, achieving significant performance gains in fluctuating wireless networks \cite{jia2024meet, yi2024athena}.
\end{itemize}

\begin{figure}[t]
    \centering
    \includegraphics[width=0.5\linewidth]{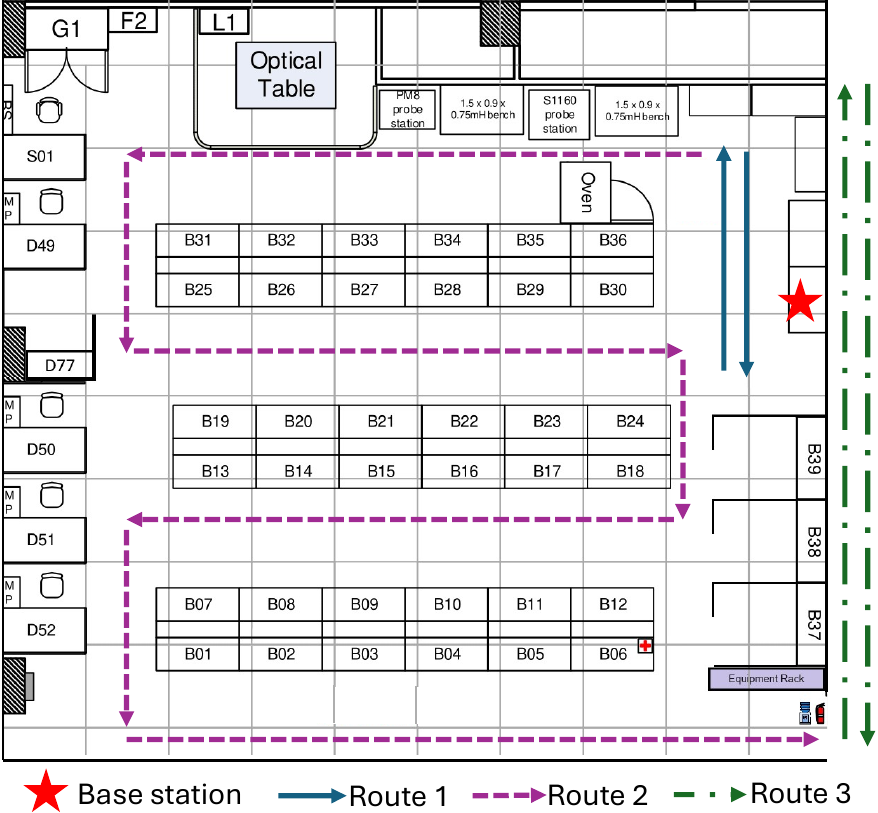}
    \caption{Lab floor plan and testing routes.}
    \label{floor plan}
\end{figure}

\textbf{\textit{Metrics.}}
We use the following metrics for the evaluation:
\begin{itemize}
    \item \textit{Network latency.} We measure the one-way delay of packets at the network layer, which serves as an indicator of network congestion.
    We keep the client and server synchronized using NTP \cite{NTP}, and capture network packets on both sides using Wireshark \cite{Wireshark}.
    The network latency is measured as the timestamp difference of the packets captured on both sides.
    \item \textit{Frame latency.} Frame latency is defined as the time interval between frame encoding at the sender and decoding at the receiver.
    A frame can only be decoded once all packets of that frame have arrived and all previously referenced frames have been decoded.
    Therefore, frame latency is a direct metric for evaluating the latency-related user experience in videos.
    We consider any frame with a delay greater than 150 ms as a video stall.
    \item \textit{Frame bitrate.} The frame bitrate arriving at the client significantly affects user experience.
    Thus, we also assess video quality based on frame bitrate.
    \item \textit{Valid bitrate.} High bitrate should only be pursued after achieving low frame delays.
    The valid bitrate refers to the average bitrate of frames with delays under the threshold of 150 ms.
\end{itemize}

Together, frame latency, stall rate, and valid bitrate comprehensively characterize the latency-throughput trade-off that governs RTC quality of experience: frame latency and stall rate measure interactivity and smoothness, while valid bitrate captures the useful throughput that contributes to perceived video quality.

\begin{figure}[t]
    \centering
    \begin{subfigure}[b]{\linewidth}
    \centering
    \includegraphics[width=\linewidth]{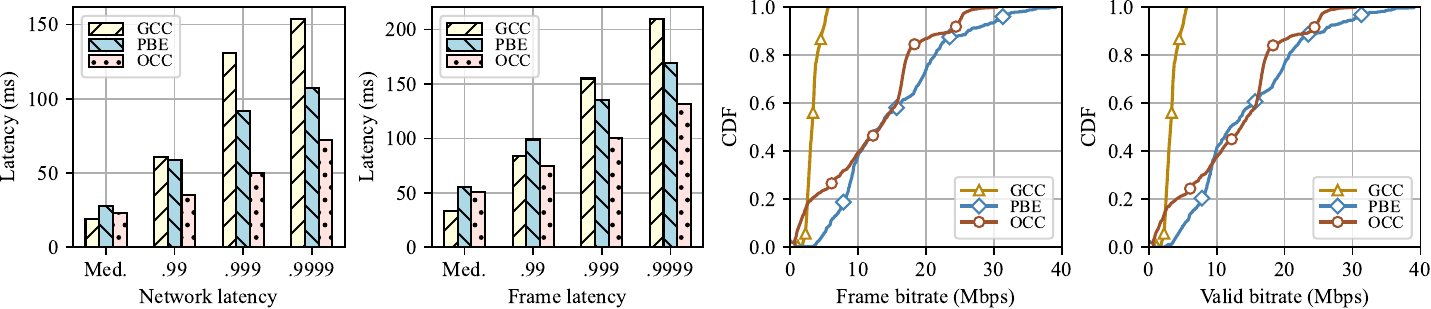}
    \caption{Evaluation results on HDR videos.}
    \label{fig0}
    \end{subfigure}
    \begin{subfigure}[b]{\linewidth}
    \includegraphics[width=\linewidth]{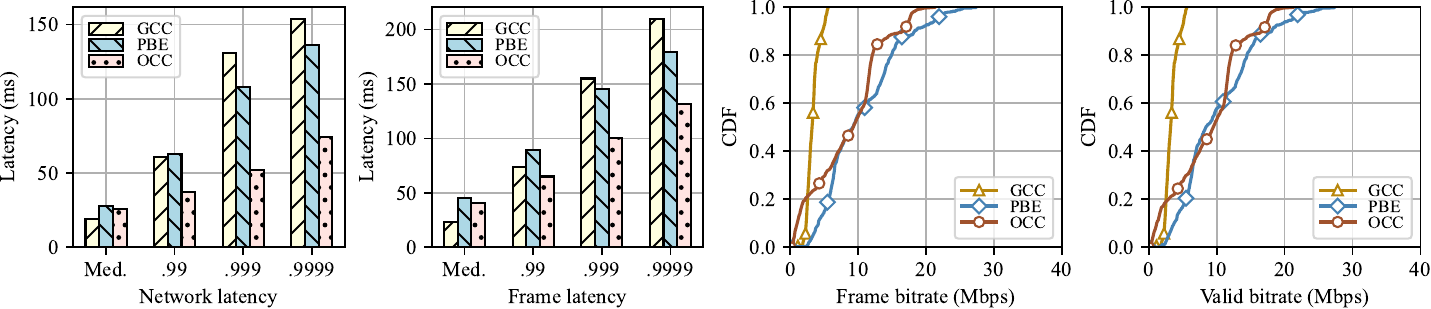}
    \caption{Evaluation results on Lecture videos.}
    \label{fig9}
    \end{subfigure}
    \caption{Overall system-level performance.}
\end{figure}

\textbf{\textit{Physical environments.}}
We test OCC in various environments, as shown in Fig. \ref{floor plan}, which are specified below:
\begin{itemize}
    \item \textit{Route 1 - Free space with good signal coverage.} The end device is positioned near the BS without any obstructions.
    \item \textit{Route 2 - Complex environment with unstable signal coverage.} The end device is moved around a laboratory setting, where signals may be blocked by test benches.
    \item \textit{Route 3 - Free space with bad signal coverage.} The end device is moved through an empty corridor outside the laboratory, where signals are obstructed by walls.
\end{itemize}

\textbf{\textit{Generality of the evaluation setup.}}
OCC exposes only a target rate to the video encoder and does not depend on codec-specific internals, making it compatible with different encoders.
We adopt x264 due to its widespread deployment in real-world RTC systems and its controllability for encoder-lag experiments.
Our over-the-air cellular testbed naturally incorporates link-layer impairments, including fading and interference. OCC targets congestion-induced delay and is orthogonal to link-layer loss recovery mechanisms, such as HARQ and RLC retransmission.

\subsection{Overall Performance Evaluation} \label{sec6.2}

As shown in Figs. \ref{fig0} and \ref{fig9}, both the latency and bitrate of the RTC flows with OCC show improvements over the baselines.
Specifically, the latency is reduced by 13\% to 68\%, while the bitrate increases by 1.2$\times$ to 3.5$\times$ across all scenarios.
These results indicate that OCC reduces the latency without sacrificing, and often improving, the video bitrate.

The comparison between the end-to-end baseline (GCC) and the physical-layer-informed baselines (PBE-CC/OCC) quantifies the benefit of explicitly obtaining ABW from the physical layer, especially under fast wireless channel fluctuations.
Moreover, the additional reduction in tail latency of OCC over PBE-CC highlights that RTC-aware temporal alignment and control are necessary to fully realize this benefit in RTC.

Moreover, we evaluate the capability of OCC to maintain consistent performance on different types of video content.
We assess the performance of OCC using both high-bitrate HDR videos (Fig. \ref{fig0}) and low-bitrate, high-frame-bursty lecture videos (Fig. \ref{fig9}).
The more pronounced RTC characteristics in lecture videos lead to a slight performance degradation for physical-layer solutions, particularly affecting PBE-CC, which does not consider RTC characteristics in its design.

\textbf{\textit{Reason for higher median latency.}}
Overall, the physical-layer-informed solutions (PBE-CC and OCC) provide significant performance gains over end-to-end solutions (GCC) in terms of tail network latency and video bitrate, although they slightly increase the median latency.
This is because PBE-CC and OCC effectively probe the available bandwidth and achieve higher sending rates, resulting in larger frame sizes and longer transmission time for each frame.
Nonetheless, the prolonged median latency remains well below the latency threshold and can be managed by configuring the video bitrate with certain limits.

\subsection{Micro Benchmarks} \label{sec6.3}

We further evaluate the performance of OCC under three RTC features: traffic burst, APP limit, and encoder lag, to individually verify the effectiveness of three designs in OCC.

\begin{figure}[t]
    \centering
    \begin{subfigure}[b]{0.49\linewidth}
        \centering
        \includegraphics[width=\linewidth]{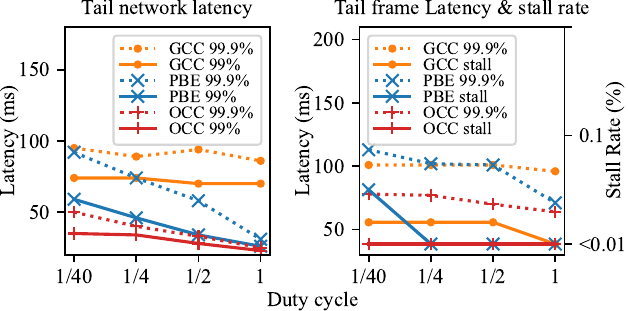}
        \caption{Latency.}
        \label{fig4}
    \end{subfigure}
    \begin{subfigure}[b]{0.24\linewidth}
        \centering
        \includegraphics[width=\linewidth]{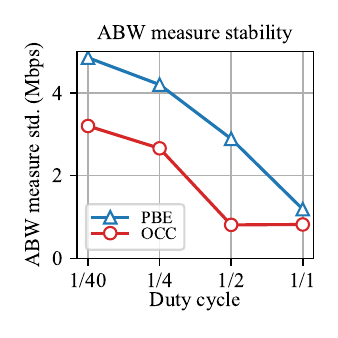}
        \caption{Measurement stability.}
        \label{fig12_1}
    \end{subfigure}
    \begin{subfigure}[b]{0.24\linewidth}
        \centering
        \includegraphics[width=\linewidth]{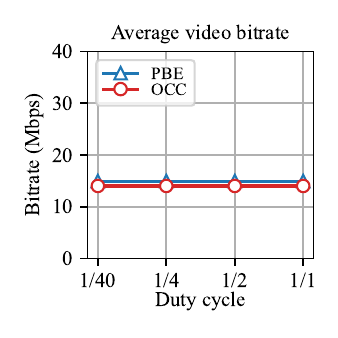}
        \caption{Average bitrate.}
        \label{fig12_2}
    \end{subfigure}
    \caption{Micro-benchmark: Frame-aware measurement.
    The duty cycle represents the fraction of one frame interval during which data packets are sent from the sender, which is simulated by adjusting the pacing rate of the RTC sender.
    A smaller duty cycle indicates a more severe traffic burst.}
\end{figure}

\textbf{\textit{Traffic burst.}}
The frame-aware bandwidth measurement enables OCC to obtain an accurate ABW estimation, which reduces tail latency by up to 41\% (Fig. \ref{fig4}) and stabilizes ABW estimation by decreasing the measurement standard deviation by up to 59\% (Fig. \ref{fig12_1}).
Notably, the more significant the traffic burst, the greater the improvements that OCC achieves, highlighting the importance of frame-aware bandwidth measurement in managing RTC's bursty traffic.
The performance of OCC converges to that of PBE-CC when the packets are fully paced (duty cycle = 1), confirming that OCC gracefully degrades to legacy physical-layer bandwidth estimation when pacing eliminates frame-level burstiness.
This demonstrates that the frame-aware measurement is a strict generalization: it improves performance under bursty or partially paced traffic while incurring no penalty under full pacing.
Moreover, the difference in video bitrate between OCC and PBE-CC is negligible (Fig. \ref{fig12_2}), indicating that the latency improvements of OCC are not due to conservative bandwidth estimation but rather to a more precise temporal granularity.

\textbf{\textit{APP limit.}}
The APP-limit-aware bandwidth estimation of OCC leads to an average bitrate increase of 31\% under APP limits (Fig. \ref{fig5}).
As the average APP limit ranges from 100\% (no limit) to 20\% (significant limit), the bitrate improvement varies from zero to 60\%. This indicates that OCC effectively guides the sender to utilize more resources when released from the APP limits.
Moreover, the negligible difference between the valid bitrate and frame bitrate in OCC suggests that the margin allocation does not introduce significant latency.

\begin{figure}[t]
    \centering
    \begin{subfigure}[b]{0.49\linewidth}
        \centering
        \includegraphics[width=\linewidth]{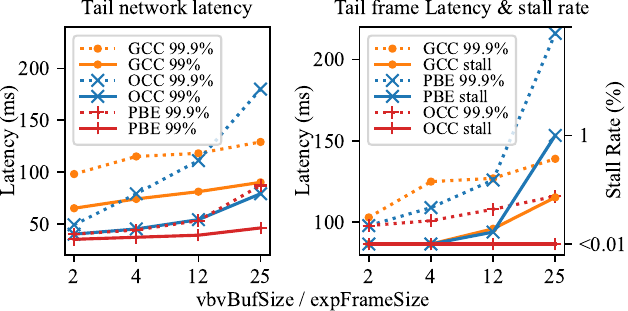}
        \caption{Latency.}
        \label{fig6}
    \end{subfigure}
    \begin{subfigure}[b]{0.245\linewidth}
        \centering
        \includegraphics[width=\linewidth]{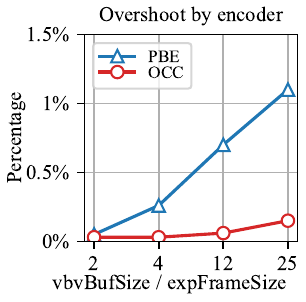}
        \caption{Encoder overshoot.}
        \label{fig11_1}
    \end{subfigure}
    \begin{subfigure}[b]{0.23\linewidth}
        \centering
        \includegraphics[width=\linewidth]{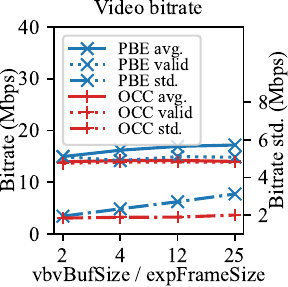}
        \caption{Average bitrate.}
        \label{fig11_2}
    \end{subfigure}
    \caption{Micro-benchmark: Encoder-friendly rate smoothing.
    We simulate different encoder lags by adjusting the VBV buffer size in the x264 codec, setting the size to 2, 4, 12, or 25 times the expected frame size of the rate control module.
    A larger VBV buffer size may result in a more severe encoder lag.}
\end{figure}

\begin{figure}[t]
    \centering
    \includegraphics[width=0.5\linewidth]{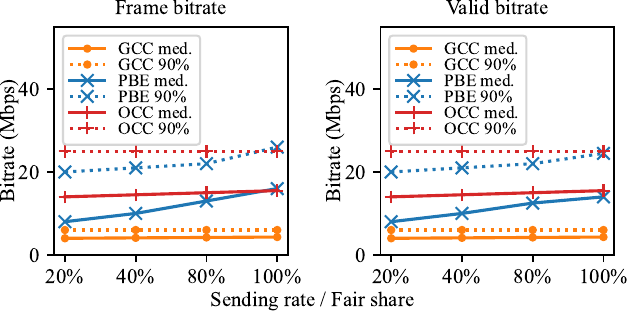}
    \caption{Micro-benchmark: APP-limit-aware bandwidth estimation.
    The ticks in x-axis indicate the percentages of actual throughput relative to the fair-share throughput in the APP limit state.
    A smaller percentage indicates a more significant APP limit.}
    \label{fig5}
\end{figure}

\textbf{\textit{Encoder lag.}}
The encoder-friendly target rate smoothing in OCC reduces the tail latency by an average of 35\%, with performance gains ranging from 13\% to 52\% as the encoder lag increases (Fig. \ref{fig6}).
This smoothing policy effectively avoids unnecessary rate increases, significantly mitigating bandwidth overshoot by the video encoder (Fig. \ref{fig11_1}).
Besides, the bitrate analysis (Fig. \ref{fig11_2}) shows that OCC does not lead to a significant drop in bitrate (with an average decrease of up to 14\% and a valid bitrate drop of 6\%) and improves bitrate consistency (reducing the standard deviation by up to 32\%).


\begin{figure}[ht]
\centering
\begin{minipage}[t]{0.66\linewidth}
    \centering
    \begin{subfigure}[t]{0.49\linewidth}
        \centering
        \includegraphics[width=\linewidth]{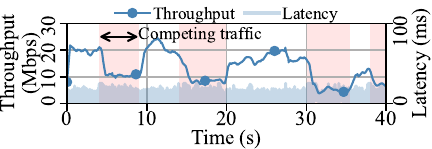}
        \caption{Coexistence with CUBIC flows.}
        \label{fig23}
    \end{subfigure}
    \hfill
    \begin{subfigure}[t]{0.49\linewidth}
        \centering
        \includegraphics[width=\linewidth]{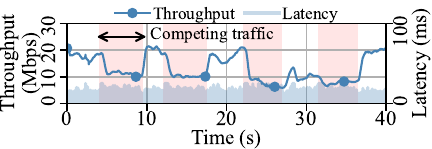}
        \caption{Coexistence with BBR flows.}
        \label{fig24}
    \end{subfigure}
    \caption{Performance of OCC when coexisting with diverse competing flows.
    The shaded areas represent the time periods when the concurrent competing traffic is active.
    OCC reacts promptly to the arrival and departure of competing flows.}
    \label{fig2324}
\end{minipage}
\hfill
\begin{minipage}[t]{0.32\linewidth}
    \centering
    \includegraphics[width=\linewidth]{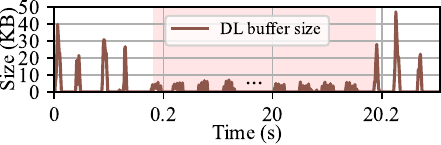}
    \captionof{figure}{Illustration of OCC's bottleneck state switching.
    The shaded area indicates when the bottleneck is in the Internet, while the remaining shows when it is in the wireless link.}
    \label{fig28}
\end{minipage}
\end{figure}

\begin{figure}[ht]
    \centering
    \begin{subfigure}[b]{0.32\linewidth}
        \centering
        \includegraphics[width=\linewidth]{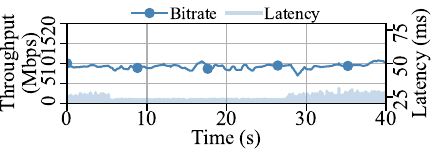}
        \caption{Static.}
        \label{fig27}
    \end{subfigure}
    \hfill
    \begin{subfigure}[b]{0.32\linewidth}
        \centering
        \includegraphics[width=\linewidth]{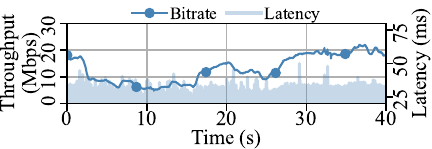}
        \caption{Move with slow walking speed.}
        \label{fig25}
    \end{subfigure}
    \hfill
    \begin{subfigure}[b]{0.32\linewidth}
        \centering
        \includegraphics[width=\linewidth]{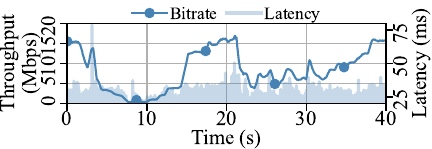}
        \caption{Move with fast walking speed.}
        \label{fig26}
    \end{subfigure}
    \caption{Performance of OCC under various levels of user mobility.
    OCC adapts quickly to ABW fluctuations caused by user mobility and keeps the latency bounded by the timely adjustment of the sending bitrate.}
    \label{fig252627}
\end{figure}

\begin{figure}[t]
\centering
\begin{minipage}[t]{0.22\linewidth}
    \centering
    \includegraphics[width=\linewidth]{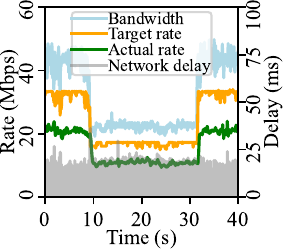}
    \captionof{figure}{Convergence test.}
    \label{fig16}
\end{minipage}
\hfill
\begin{minipage}[t]{0.22\linewidth}
    \centering
    \includegraphics[width=\linewidth]{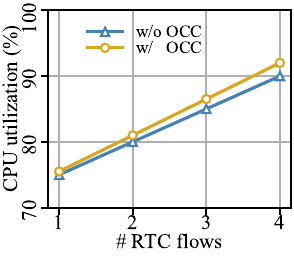}
    \captionof{figure}{CPU overhead.}
    \label{fig13}
\end{minipage}
\hfill
\begin{minipage}[t]{0.22\linewidth}
    \centering
    \includegraphics[width=\linewidth]{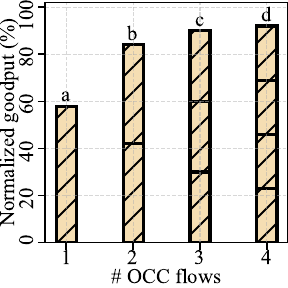}
    \captionof{figure}{Internal fairness.}
    \label{fig18}
\end{minipage}
\hfill
\begin{minipage}[t]{0.22\linewidth}
    \centering
    \includegraphics[width=\linewidth]{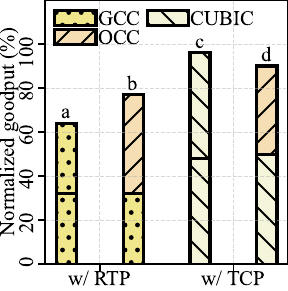}
    \captionof{figure}{External fairness.}
    \label{fig17}
\end{minipage}
\end{figure}

\subsection{Dynamic Scenarios} \label{sec6.4}

Moreover, we evaluate the performance of OCC under dynamic network conditions, including multi-flow coexistence, bottleneck state switching, and varying user mobility.

\textbf{\textit{Coexistence with diverse competing flows.}}
We explore the performance of OCC when coexisting with diverse competing flows from other users who share the same cellular BS, as shown in Fig. \ref{fig2324}.
OCC dynamically adjusts its sending rate in response to competing flows, rapidly acquiring available capacity while maintaining stable latency.
One drawback of OCC is that it is not applicable when multiple flows towards the same device are competing for bandwidth, but this case is relatively rare because the RTC flow is usually dominant.

\textbf{\textit{Bottleneck state switching.}}
We emulate Internet bottleneck by limiting the egress bandwidth at the video server using Linux \textit{\textbf{tc}} \cite{tc}.
Fig. \ref{fig28} illustrates OCC's behavior during bottleneck switches.
By monitoring the packet arrival pattern in the DL buffer at the cellular BS, OCC accurately detects whether the bottleneck lies within the wireless link or the Internet.
Consequently, OCC seamlessly falls back to GCC when the bottleneck shifts to the Internet, and regains control when the wireless link becomes the bottleneck again.

\textbf{\textit{Adaptation to various levels of user mobility.}}
We demonstrate how OCC behaves under various levels of mobility.
As shown in Fig. \ref{fig252627}, OCC effectively bounds latency by adapting quickly to changes in channel conditions caused by user mobility, even at a fast walking speed.
Although transient latency spikes occur when the user moves abruptly into areas with poor signal coverage (around $t=4$ s and $t=22$ s in Fig. \ref{fig26}), the latency remains below the threshold and quickly recovers due to timely and adequate rate reduction.

\subsection{OCC Deep Dive} \label{sec6.5}
Finally, we report the convergence speed, fairness, and runtime overhead of OCC.

\textbf{\textit{Convergence speed.}}
With explicit knowledge of ABW through physical-layer information, OCC converges to bandwidth changes within $1$ s (Fig. \ref{fig16}).
While the target rate immediately reflects the ABW, the overall convergence time also depends on the encoder lag (see Section \ref{encoder lag}).

\textbf{\textit{CPU overhead.}}
We measure the CPU utilization of OCC on the O-RAN testbed with different numbers of concurrent RTC flows (Fig. \ref{fig13}).
The baseband processing of O-RAN consumes a significant amount of CPU resources ($\sim$ 75\% with one RTC flow), and the processing required for OCC logic does not impose a substantial computational burden on the BS's CPU.

\textbf{\textit{Internal fairness.}}
We analyze whether OCC affects the bitrate fairness in a steady state when optimizing multiple RTC flows simultaneously.
The bars $a$ to $d$ in Fig. \ref{fig18} illustrate the goodput of one to four flows using OCC on the same BS.
We observe that the bitrate fairness among multiple OCC flows is effectively maintained under steady-state conditions.

\textbf{\textit{External fairness.}}
When competing with both GCC and CUBIC, the performance improvement of OCC does not come at the cost of sacrificing the performance of other flows.
We measure the bitrate of two competing flows: one optimized by OCC and the other either a legacy RTC flow (GCC) or a TCP flow (CUBIC), as presented in Fig. \ref{fig17}.

\section{Related Work}
\textbf{\textit{Physical-layer-informed congestion control.}}
piStream \cite{xie2015pistream} harnesses the physical-layer information for dynamic adaptive video streaming (DASH) over LTE and demonstrates performance gains in video bitrate and reduced stall rate compared to state-of-the-art DASH adaptation protocols.
Similarly, CLAW \cite{xie2017accelerating} utilizes the physical-layer information to probe the maximum available bandwidth, thereby reducing the loading time of the web page.
PBE-CC \cite{xie2020pbe} integrates fine-grained bandwidth measurements from the physical layer into the TCP congestion control, achieving higher average throughput while reducing tail latency.
However, these solutions do not consider the characteristics of RTC, resulting in suboptimal RTC performance.
PERCEIVE \cite{lee2020perceive} focuses on predicting cellular uplink bandwidth from scheduling patterns, while OCC explicitly aims to obtain the physical-layer capacity and utilize it for closed-loop RTC rate control by addressing RTC-specific challenges, including traffic bursts, APP limits, and encoder lag.

\textbf{\textit{Application-aware physical-layer design.}}
Tutti \cite{xu2022tutti} enforces a deadline-sensitive PRB allocation in 5G RAN for latency-critical video analytics applications, resulting in improved quality of experience (QoE) for video analytics.
Zipper \cite{balasingam2024application} provides application-level service assurance through 5G RAN slicing, allocating minimal radio resources that meet the service level agreement (SLA) of all APPs within each slice.
However, the above solutions require modifications to the operational logic of the cellular BS, which requires careful justification on their necessity and practicality.

\section{Conclusion}
We propose OCC, a cross-layer solution for RTC over mobile networks.
By leveraging the physical-layer information, OCC addresses the latency-bitrate trade-off in dynamic wireless environments.
OCC employs a frame-aware measurement approach to obtain a reliable ABW estimation under RTC's bursty traffic.
Moreover, it effectively probes the available bandwidth in the APP limit state and adjusts the target rate in an encoder-friendly manner.
We implement OCC in an O-RAN testbed as a proof-of-concept and conduct extensive over-the-air performance evaluations.
Experiments show that OCC reduces network latency by 13\% to 68\%, while improving video frame bitrate by $1.2\times$ to $3.5\times$ across different scenarios.

\bibliographystyle{plainnat}
\bibliography{main_arxiv}

\end{document}